\documentclass[preprintnumbers, floatfix, preprintnumbers, letterpaper, superscriptaddress,nofootinbib, twocolumn]{revtex4}
\pdfoutput=1
\usepackage{graphicx}
\usepackage{microtype}
\usepackage{amsmath}
\usepackage{amssymb}
\usepackage{subfigure}
\usepackage{hyperref}
\usepackage{url}
\usepackage{xcolor}
\usepackage{color}
\usepackage{mathrsfs}
\usepackage{calrsfs}
\usepackage{amsfonts}
\usepackage{latexsym}
\usepackage{ragged2e}
\usepackage{epsfig}
\usepackage{textcomp}
\usepackage{phaistos}
\makeatletter
\renewcommand\@makefnmark{\hbox{\@textsuperscript{\normalfont\color{purple}\@thefnmark}}}
\renewcommand\@makefntext[1]{%
  \parindent 1em\noindent
            \hb@xt@1.8em{%
                \hss\@textsuperscript{\normalfont\@thefnmark}}#1}
\makeatother

\usepackage{caption}
\DeclareCaptionJustification{justified}{\leftskip=0pt \rightskip=0pt \parfillskip=0pt plus 1fil}
\definecolor{vividviolet}{rgb}{0.62, 0.0, 1.0}
\definecolor{amaranth}{rgb}{0.9, 0.17, 0.31}
\definecolor{palatinateblue}{rgb}{0.15, 0.23, 0.89}
\definecolor{brightpink}{rgb}{1.0, 0.0, 0.5}
\definecolor{cornflowerblue}{rgb}{0.39, 0.58, 0.93}
\definecolor{deepcarminepink}{rgb}{0.94, 0.19, 0.22}
\definecolor{radicalred}{rgb}{1.0, 0.21, 0.37}

\hypersetup{ linktoc=all,
    colorlinks, linkcolor={palatinateblue},
    citecolor={brightpink}, urlcolor={amaranth}
}

\graphicspath{{Images/}}

\graphicspath{{Images/}}



\def\sideremark#1{\ifvmode\leavevmode\fi\vadjust{\vbox to0pt{\vss
 \hbox to 0pt{\hskip\hsize\hskip1em
 \vbox{\hsize1.5cm\tiny\raggedright\pretolerance10000
 \noindent #1\hfill}\hss}\vbox to8pt{\vfil}\vss}}}%
\begin{document}

\title{Entropy production and correlation spreading in the interaction between particle detector and thermal baths}

\author{Hao Xu}
\email{haoxu@yzu.edu.cn}
\affiliation{Center for Gravitation and Cosmology, College of Physical Science and Technology, Yangzhou University, \\180 Siwangting Road, Yangzhou City, Jiangsu Province  225002, China}
\affiliation{Shanghai Frontier Science Center for Gravitational Wave Detection, Shanghai Jiao Tong University, Shanghai 200240, China}

\author{Si Yu Chen}
\affiliation{Center for Gravitation and Cosmology, College of Physical Science and Technology, Yangzhou University, \\180 Siwangting Road, Yangzhou City, Jiangsu Province  225002, China}

\begin{abstract}
Entropy production is the key to the second law of thermodynamics, and it is well defined by considering a joint unitary evolution of a system $S$ and a thermal environment $E$. However, due to the diversity of the initial state and Hamiltonian of the system and environment, it is hard to evaluate the characterisation of entropy production. In the present work, we propose that the evolution of $S$ and $E$ can be solved non-perturbatively in the framework of Gaussian quantum mechanics (GQM). We study the entropy production and correlation spreading in the interaction between Unruh-DeWitt-like particle detector and thermal baths, where the particle detector is set to be a harmonic oscillator and the thermal baths are made of interacting and noninteracting Gaussian states. We can observe that the entropy production implies quantum recurrence and shows periodicity. In the case of interacting bath, the correlation propagates in a periodic system and leads to a revival of the initial state. Our analysis can be extended to any other models in the framework of GQM, and it may also shed some light on the AdS/CFT correspondence.
\end{abstract}

\maketitle

\emph{Introduction}.---Entropy production, which is non-negative and becomes zero only in the case where the thermodynamic process is reversible, serves as the key to the second law of thermodynamics. It predicts there is a lack of time reversal in the irreversibility thermodynamic process coming from the time arrow \cite{Batalhao2015}. Its research has also become one of the most important topics in modern physics \cite{Landi:2020bsq}.

However, the exact definition of entropy production was, surprisingly, not universal. The natural candidate for the definition of entropy, namely, the von Neumann entropy, is invariant under unitary evolution, thus it can not reflect the irreversibility in the evolution of the system. Studies on entropy production were often based on specific models and different assumptions, until it was addressed in \cite{esposito2010} by considering a joint unitary evolution of a system $S$ and an environment $E$ under specific conditions. A general formula of entropy production was defined, and this method was also later used to clarify the long controversial Landauer's principle in \cite{Reeb2013}. The setup for the joint evolution was based on four assumptions: (i) both the system $S$ and environment $E$ are described by Hilbert spaces, (ii) the environment is initially in a thermal state $\rho_E=e^{-\beta \hat{H}_E}/\text{Tr}(e^{-\beta \hat{H}_E})$, where $\hat{H}_E$ is the Hamiltonian of the environment and $\beta$ is the inverse temperature, (iii) system and environment are uncorrelated initially, $\rho_{SE}=\rho_{S}\otimes\rho_{E}$, (iv) the process proceeds by unitary evolution $\rho'_{SE}=U\rho_{SE}U^{\dagger}$. If all the four assumptions are satisfied, Landauer's principle holds and the entropy production $\zeta$ can be defined as
\begin{equation}
\zeta= I(S':E')+D(\rho'_E||\rho_E),
\end{equation}
where the $I(S':E'):=S(\rho'_S)+S(\rho'_E)-S(\rho'_{SE})$ is the mutual information (MI) that measures the correlation between the system and environment, while the $S(\bullet)$ denotes the von Neumann entropy and $\rho'_{S/E}:=\text{Tr}_{E/S}[\rho'_{SE}]$ is reduced density matrix. The $D(\rho'_E||\rho_E):=\text{Tr}(\rho'_E\ln{\rho'_E})-\text{Tr}(\rho'_E\ln{\rho_E})$ is the relative entropy measures the displacement of the environment from the initial thermal equilibrium state. The non-negativity of entropy production follows from the fact that both $I(S':E')$ and  $D(\rho'_E||\rho_E)$ are non-negative.

On the other hand, although the setup in \cite{esposito2010,Reeb2013} are general, it is also minimal thus can not provide further information about the thermodynamical process. It is natural for us to ask how these two terms (MI and relative entropy) contribute to entropy production. It was often held that, compared with the MI, the relative entropy is negligible for large thermal reservoirs \cite{li2017, strasberg2017, chen2017, engelhardt2018, manzano2018, you2018, li2019, santos2019, bera2019}. However, recently \cite{Ptaszynski2019} it was found that the MI is strongly bounded from above by the Araki-Lieb inequality \cite{araki1970}. The entropy production could be time-extensive, so the entropy production could also predominantly come from the relative entropy that measures the displacement of the environment from equilibrium state $\rho_E$. In \cite{Ptaszynski2019} a numerical analysis of fermions with noninteracting resonant levels was also presented to support this point.

Although in some cases the relative entropy starts to dominate after the early stage of evolution, this does not mean we have fully understood the characterization and assessment of entropy production. Ultimately, the details of entropy production are determined by the initial state and the Hamiltonian of the total system. The initial state only needs to satisfy assumptions (i)-(iii), and there are no restrictions of the Hamiltonian. In principle, there could be infinite kinds of choices. To make matters worse, the Hamiltonian of $S$ may be very different from $E$, making it almost impossible to solve the evolution of the joint system. In many cases, we can only use perturbative method to solve the equation order by order. 

However, perturbative methods naturally have disadvantages. Long time interaction, strong coupling, or high-average-energy exchange processes will make the perturbative expansion invalid. One way to address these issues is to impose a few restrictions on the choice of Hamiltonian and the initial state of the total system, so that the time evolution can be solved non-perturbatively. This can be achieved by using the symplectic formalism in Gaussian quantum mechanics (GQM), which maps the Schr\"{o}dinger equation in the infinite-dimensional Hilbert space onto a linear evolution of finite-dimensional phase space \cite{Adesso2007,Adesso2014}. GQM provides a way for us to solve a class of models completely. It has been widely used in quantum optics \cite{Olivares2012}, and recently it is also applied to analogue gravity systems \cite{Bruschi2013a}, relativistic quantum field theory (QFT) \cite{Bruschi2013,Brown2012,Bruschi2015,Bruschi2020} and quantum thermodynamics \cite{Kerstjens2018}. Using GQM, we can solve the time evolution of the total system and gain access to the states at any moment of time for finite-size, finite-time, and strong-coupling regimes. 

In the present work we will apply GQM to study the entropy production and correlation spreading in the interaction between Unruh-DeWitt-like particle detector and thermal baths. The traditional Unruh-DeWitt detector, which is often described by an accelerating qubit coupled to the vaccum state of QFT \cite{DeWitt1979}, is used to characterize the Unruh effect \cite{Unruh1976,Unruh1983}. However, in our case we set the particle detector to be an inertial harmonic oscillator, which represents the system $S$ in the interaction. The environments $E$ made of interacting or noninteracting baths are described by the thermal Gaussian states. The joint evolution of $S$ and $E$ can be solved in the framework of GQM. Since the $S$ is a detector, the interaction can also be viewed as a measurement process. In principle, we can obtain the information of $E$ by studying the feedback obtained by $S$. In the present work we concentrate on the interaction and correlation between the particle detector and thermal states. 

Our results show that the entropy production is not dominated by the relative entropy as in \cite{Ptaszynski2019} when considering the long-time interactions, but implies quantum recurrence and shows periodicity. The statement of entropy production predominantly comes from relative entropy is only part of the story, and there is still much to study about the evolution of the whole system. In the case of interacting bath, the correlation propagates in a periodic system and lead to a revival of the initial state. Similar examples of this phenomenon can also be found in AdS/CFT (Anti-de Sitter/Conformal Field Theory) correspondence \cite{Maldacena:1997re,Witten:1998qj}, where the evolution of CFT after a quantum quench can be explained by the left and right-moving quasiparticles \cite{Cardy:2014rqa}. Henceforth we adopt the natural unit system, setting $c=\hbar=k_B=1$. \\

\emph{Gaussian quantum mechanics}.--- In this section we give a brief review of the GQM. The reader is referred to \cite{Adesso2007,Adesso2014} for a detailed introduction to this topic. For a bosonic system with $N$ modes, we can form a vector of operators as
\begin{equation}
\boldsymbol x=(\hat{q}_1,\hat{p}_1,\cdots,\hat{q}_N,\hat{p}_N)^{{T}},
\end{equation}
where the canonical quadrature operators $\hat{q}_i=(\hat{a}_i+\hat{a}_{i}^{\dag})/\sqrt{2}$ and $\hat{p}_i=i(\hat{a}_{i}^{\dag}-\hat{a}_i)/\sqrt{2}$ satisfy $[\hat{q}_i,\hat{p}_i]=i\delta_{ij}$. The phase space formed by the above vector is a symplectic space. In GQM we only consider the Gaussian states that are fully described by the first and second moments of their quadratures, and the Gaussian states stay Gaussian under the evolution with any time-dependent \emph{quadratic} Hamiltonian. For simplicity we set the first moment of the vector to be zero and characterize the state of the system as the $2N \times 2N$ covariance matrix $ \boldsymbol \sigma$ whose entries are given by
\begin{equation}
\sigma_{ab}=\langle \hat{x}_a\hat{x}_b+\hat{x}_b \hat{x}_a\rangle=\text{Tr}[\rho(\hat{x}_a \hat{x}_b+\hat{x}_b \hat{x}_a)].
\end{equation}
Any combined state of two systems $A$ and $B$ has the form
\begin{equation}
\sigma_{AB}=\begin{pmatrix} 

    \boldsymbol {\sigma}_A & \boldsymbol{\gamma}_{AB} \\

    \boldsymbol{\gamma}_{AB}^T & \boldsymbol{\sigma}_B

\end{pmatrix} ,
\end{equation}
where $\boldsymbol{\sigma}_A$ and $\boldsymbol{\sigma}_B$ are the states of $A$ and $B$ respectively, and the $\boldsymbol{\gamma}_{AB}$ represents the correlation between them. For systems under unitary transformation $U$ generated by a time-dependent quadratic Hamiltonian $H(t)$, it corresponds to a linear symplectic transformation on the phase space: $\boldsymbol x\rightarrow \boldsymbol x'= U^{\dag}\boldsymbol x U=S\boldsymbol x$. Similarly the covariance matrix transforms as $\boldsymbol {\sigma}\rightarrow \boldsymbol {\sigma}'=S \boldsymbol {\sigma} S^T$.

Since $H(t)$ is quadratic, we can write it as $H(t)=\boldsymbol x^T F(t) \boldsymbol x$, where $F(t)$ is a $2N \times 2N$ phase space matrix. The symplectic evolution matrix $S(t)$ generated by this Hamiltonian obeys a Schr\"{o}dinger-like equation:
\begin{equation}
\frac{dS(t)}{dt}=\Omega F_s(t)S(t),
\label{equation}
\end{equation}
where $F_s=F+F^T$ and $\Omega:=\bigoplus^N_{i=1}\begin{pmatrix} 

    0 & 1 \\

    -1 & 0

\end{pmatrix} $ is the  symplectic form. Once we have the formula of the total Hamiltonian, we can solve the above differential Schr\"{o}dinger-like equation numerically. The inital condition of the equation is $S(0)=\openone$. We can decompose the continuous time $T$ into many steps and solve the equation step by step. An example of the codes, both in Matlab and Python, which are developed by the authors of \cite{Kerstjens2018}, are available in \cite{Kerstjens2018a}.

In quantum statistical physics we always need to do the tracing over $\rho$ to obtain the expectation value of operator $\hat{O}$ as $O=\text{Tr}[\rho \hat{O})]$. In GQM we only care about the first and second moments, and the information of the states has been encoded in the covariance matrix, so we can always represent the expectation values of other physical quantities as the function of covariance matrix. For example, the average energy reads $\langle H \rangle=\frac{1}{2}\text{Tr}[F\sigma]$. If we diagonalize the system's Hamiltonian as $F=\frac{1}{2}\text{diag}(\omega_1,\omega_1,\omega_2,\omega_2, \ldots)$, the thermal state will be
\begin{equation}
\sigma_T=\bigoplus^N_{i=1}\begin{pmatrix} 

    \nu^{(th)}_i & 0 \\

    0 &  \nu^{(th)}_i

\end{pmatrix}, \qquad \nu^{(th)}_i=\frac{e^{\omega_i/T}+1}{e^{\omega_i/T}-1},
\label{thermal}
\end{equation}
which corresponds to a tensor product state with a diagonal density matrix given by
\begin{equation}
\bigotimes_i \frac{2}{\nu^{(th)}_i+1}\sum_{n_i=0}^{\infty}\left( \frac{\nu^{(th)}_i-1}{\nu^{(th)}_i+1}\right) |n_i\rangle \langle n_i|,\label{density}
\end{equation}
where $|n_i\rangle$ denotes the number state in the Fock space. Each mode with frequency $\omega_i$ is a Gaussian state in thermal equilibrium at a temperature $T$, and the average photons number
\begin{equation}
\bar{n}_i=\frac{\nu^{(th)}_i-1}{2}=\frac{1}{e^{\omega_i/T}-1}.
\end{equation}
Since we already obtain the density matrix \eqref{density}, we can calculate the the von Neumann entropy directly. Finally we have
\begin{equation}
S(\sigma)=\sum^N_{i=1}f(\nu_i^{(th)}),
\end{equation}
where
\begin{equation}
f(\nu)=\frac{\nu+1}{2}\log{\frac{\nu+1}{2}}-\frac{\nu-1}{2}\log{\frac{\nu-1}{2}}.
\end{equation}
Of course, not all Gaussian states are thermal, nevertheless, there always exists a symplectic matrix to diagonalize any Gaussian state. Given any joint state of $S$ and $E$, we can calculate the von Neumann entropy directly, thus obtaining the MI \cite{Demarie2012}. On the other hand, from \cite{Reeb2013} we know the entropy production can also be written as $\zeta=\beta\left(H(\rho'_E)-H(\rho_E)\right)-(S(\rho_S)-S(\rho'_S))$, evaluating the energy change of $E$ and the entropy change of $S$, we can also obtain the relative entropy $D(\rho'_E||\rho_E)$. 

\emph{Models}.--- First we choose a harmonic oscillator with frequency $\omega$ to be the particle detector (system $S$), and we set a one-dimensional free massless scalar QFT in a \emph{cavity} to be the environment $E$ so that we can obtain discrete modes by applying an infrared cutoff and Dirichlet boundary condition. The Hamiltonian of $E$ reads $\hat{H}_{E}=\sum_{j=1}^{N}\omega_j \hat{a}^{\dag}_j \hat{a}_j$, where $N$ is the number of modes we include, $\omega_j=j\pi/L$ and $L$ is the scale of the cavity. The interaction Hamiltonian $\hat{H}_{\text{int}}(t)=\lambda \chi(t)\mu \phi[x]$, in which $\lambda$ is coupling constant, $\chi(t)$ is the switching function that controls the interaction, $\mu=\hat{a}_s+\hat{a}_s^{\dag}$ is the monopole moment of the particle detector, and $\phi[x] = \sum_{j=1}^{N}\left( \hat{a}_j u_j\left[x\right]+\hat{a}^{\dagger}_ju_j^*\left[x\right] \right)$ is the field operator at the position of the particle detector in the cavity, where $u_j(x)=\sin(j\pi x)$ is the basis and the normalization constant is absorbed in the coupling constant. This formula of $\hat{H}_{\text{int}}$ means the particle detector interacts with all modes of the environment. 

Note that the above Hamiltonian is written in the form of creation-annihilation operators, not position-momentum operators as in $\boldsymbol x$. Nevertheless, we can always do the transformation to obtain the right form of $F(t)$.

Since we are considering non-interacting cavity QFT, the $F$ for $\hat{H}_E$ is naturally diagonalized. If at $t=0$ the initial state the particle detector is in temperature $T_S$ and environment is in temperature $T_E$, the covariance matrix can be written as the direct sum of \eqref{thermal}. Setting the smooth and compactly-supported switching function as \cite{Kerstjens2018}
\begin{equation}
    \chi(t)=\begin{cases}
    0 & t < 0 \\
    \frac{1}{2}-\frac{1}{2}\tanh\cot\frac{\pi t}{\delta} & 0\leq t < \delta\\
    1 & \delta \leq t < \tau-\delta \\
    \frac{1}{2} + \frac{1}{2}\tanh\cot\frac{\pi (t-\tau)}{\delta} & \tau-\delta \leq t <\tau \\
    0 & t > \tau
    \end{cases},
    \label{swaping}
\end{equation}
we can solve the eq.\eqref{equation} numerically and obtain $S(t)$. 

\begin{figure}
\begin{center}
\includegraphics[width=0.45\textwidth]{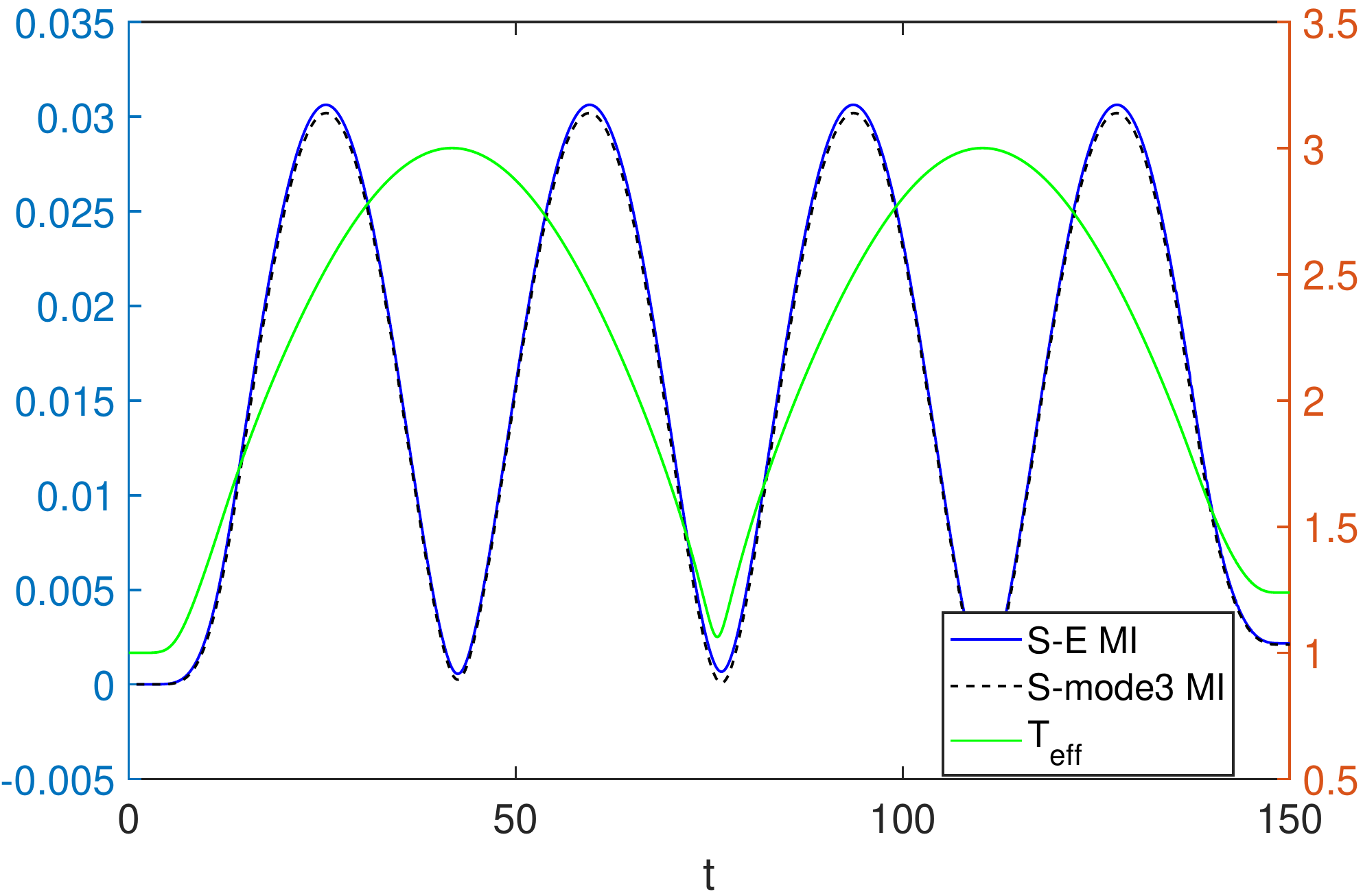}
\caption{Evolution during the interaction between $S$ and $E$ for the cavity QFT. We set $N=30$, $L=1$, $T_S=1$, $T_E=3$, $\lambda=0.05$, $\delta=0.1\tau$, $\tau=150$, and the frequency of the system $\omega=3\pi$. The green solid line is the effective temperature of $S$, corresponding to the right vertical axis, while the other curves are the MI, corresponding to the left vertical axis: the blue solid line is the MI between $S$ and $E$, and the black dashed line is the MI between $S$ and mode 3 of the cavity QFT.}
\label{fig1}
\end{center}
\end{figure}

\begin{figure}
\begin{center}
\includegraphics[width=0.45\textwidth]{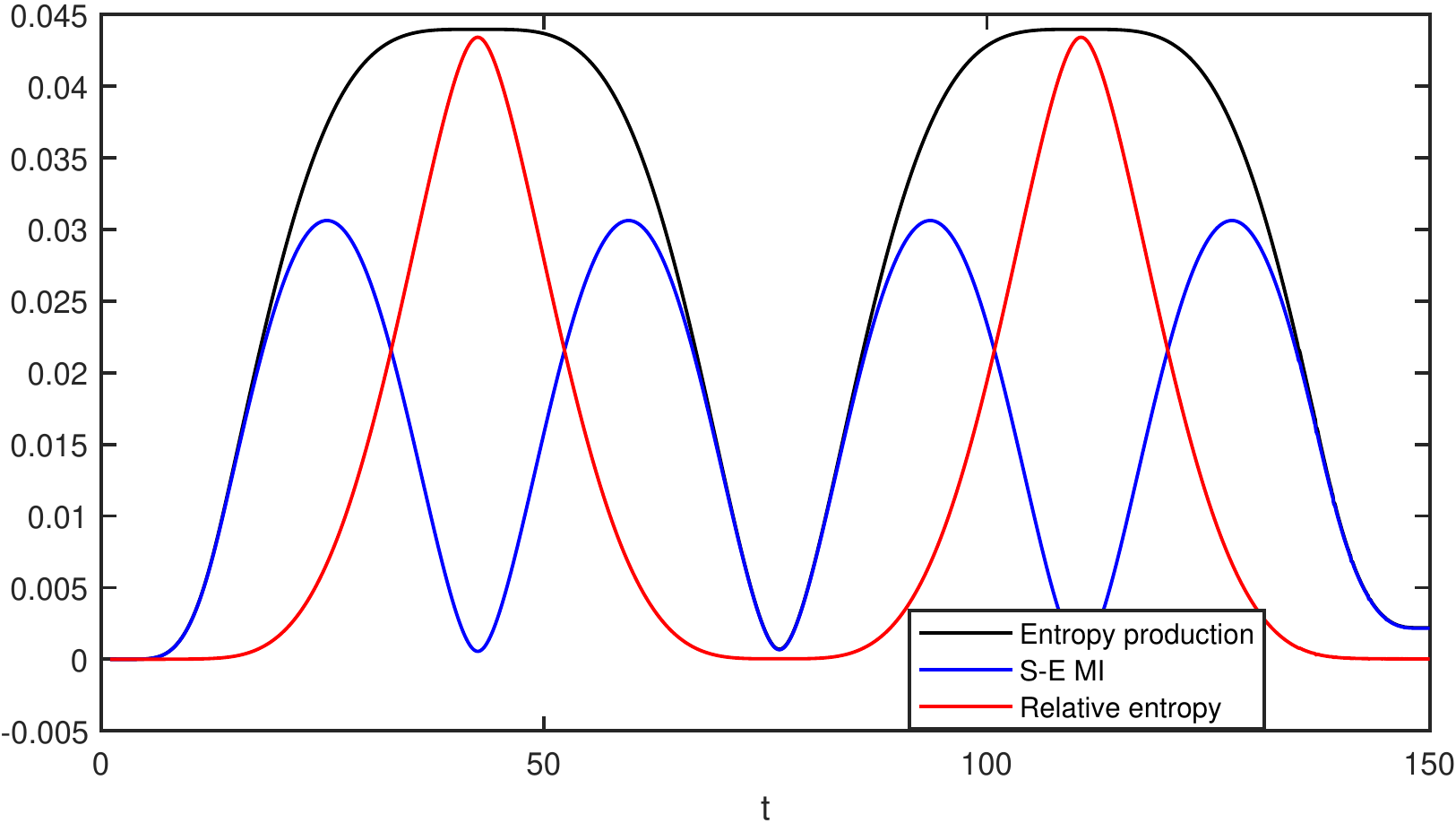}
\caption{Evolution during the interaction between $S$ and $E$ for the cavity QFT. The parameters are set as in the FIG.\ref{fig1}. The black solid line is the entropy production. The blue solid line the MI between $S$ and $E$, and the red solid line is the relative entropy $D(\rho'_E||\rho_E)$.}
\label{fig2}
\end{center}
\end{figure}

In FIG.\ref{fig1} and FIG.\ref{fig2} we present some numerical results. We set $N=30$, $L=1$, $T_S=1$, $T_E=3$, $\lambda=0.05$, $\delta=0.1\tau$, $\tau=150$. The authors of \cite{Kerstjens2018} refer to $\delta$ as the ramp-up time of the interaction. Note that the switching on and off could not be too fast ($\delta$ could not be too small), otherwise the transition rate of the system $S$ would encounter divergence \cite{Sriramkumar:1994pb,Louko:2006zv}. The frequency $\omega$ of the system $S$ is $3\pi$, which is equal to the third modes $\omega_3$ (mode 3) of the cavity QFT. The numerical result shows that the convariance matrix of the system $S$ effectively has the form of $\begin{pmatrix} 

    \nu & 0 \\

    0 &  \nu

\end{pmatrix}$, thus according to \eqref{thermal} we can define the effective temperature of the mode as $T_{\text{eff}}=\omega \ln{\frac{\nu-1}{\nu+1}}$. In FIG.\ref{fig1} we find the effective temperature of $S$ (green solid line), does not reach a steady state of thermal equilibrium as one would expect. The effective temperature of $S$, and also the MI between $S$ and $E$ (blue solid line), are both oscillating. This MI is almost indistinguishable from the one between the $S$ and the mode 3 (black dashed line). This is consistent with the result obtained by using the perturbative method that indicates the system can only be affected by a small neighbourhood of frequency $\omega$ of the cavity QFT \cite{Xu2021}. An analogous phenomenon in classical mechanics was reported in \cite{Smith2008}.

In FIG.\ref{fig2} we study the entropy production $\zeta$, $I(S':E')$ and $D(\rho'_E||\rho_E)$. Similar to FIG.\ref{fig1}, these three quantities are all oscillating. In the early stage of evolution, entropy production mainly comes from MI. After that, MI decays and the relative entropy increases, meaning the displacement of the environment from equilibrium begins to become significant while the correlation between $S$ and $E$ decreases. After reaching the peak, the relative entropy decreases rapidly, and entropy production also begins to decline soon. Meanwhile, the effective temperature is also decreasing. Eventually they all reach their lowest point and then a new cycle begins. Our results show that neither MI nor relative entropy has always provided a dominant contribution to the entropy production throughout the evolutionary process. They are each becoming dominant terms at different times.

From the above analysis we can conclude that although the cavity QFT contains various modes, the interaction is largely determined by the mode with the same frequency of $S$. The effective temperature of the system, and also the  $I(S':E')$ and $D(\rho'_E||\rho_E)$, are all oscillating and implying quantum recurrence.

Next, we consider the environment $E$ in which all modes share the same frequency with the system $S$, but there is an interaction between adjacent modes of the environment. This model is also used as the heat bath of quantum Otto engine in quantum thermodynamics \cite{Kerstjens2018}. In continuous and massless limit it also reduces to the CFT with central charge $c=1$ \cite{Chapman:2018hou,Camargo:2020yfv}. We not only want to know how the correlation between system and environment evolves, but also how it propagates in the environment. The Hamiltonian of the environment reads
\begin{equation}
    \hat{H}_E=\sum_{i=1}^{N}\frac{\omega}{2}\left(\hat{p}_{i}^2+\hat{q}_{i}^2\right) + \sum_{i=1}^N\alpha \hat{q}_{i}\hat{q}_{i+1},
    \label{Hbath}
\end{equation}
where $N$ is the number of modes, $\omega$ is the frequency, and $\alpha$ is the position-position coupling constant. Applying the periodic boundary condition $\hat{q}_{N+1}=\hat{q}_1$, $E$ becomes an oscillator chain. 

The interaction Hamiltonian can be set as
\begin{equation}
    \hat{H}_{\text{int}}=\lambda \chi(t) \hat{q}_s \sum_{i \in \{\text{int}\}} \hat{q}_i 
    \label{Hint}
\end{equation}
where $\lambda$ and $\chi(t)$ are still the coupling constant and switching function respectively, $q_s$ is the position operator of system $S$, and \{int\} denotes the modes that $S$ interacts with. For simplicity we only consider $S$ interacts with one mode of the $E$, and without loss of generality, we set $i=1$.

Note that now due to the position-position coupling constant $\alpha$, the phase space matrix $F$ corresponding to the $\hat{H}_E$ is no longer diagonal. The corresponding covariance matrix of the thermal state with temperature $T_E$ is also not diagonal. Nevertheless, we can symplectically diagonalize the phase space matrix to obtain the corresponding diagonal covariance matrix and then transform it back to the physical basis to find the thermal state of $E$.

\begin{figure}
\begin{center}
\includegraphics[width=0.45\textwidth]{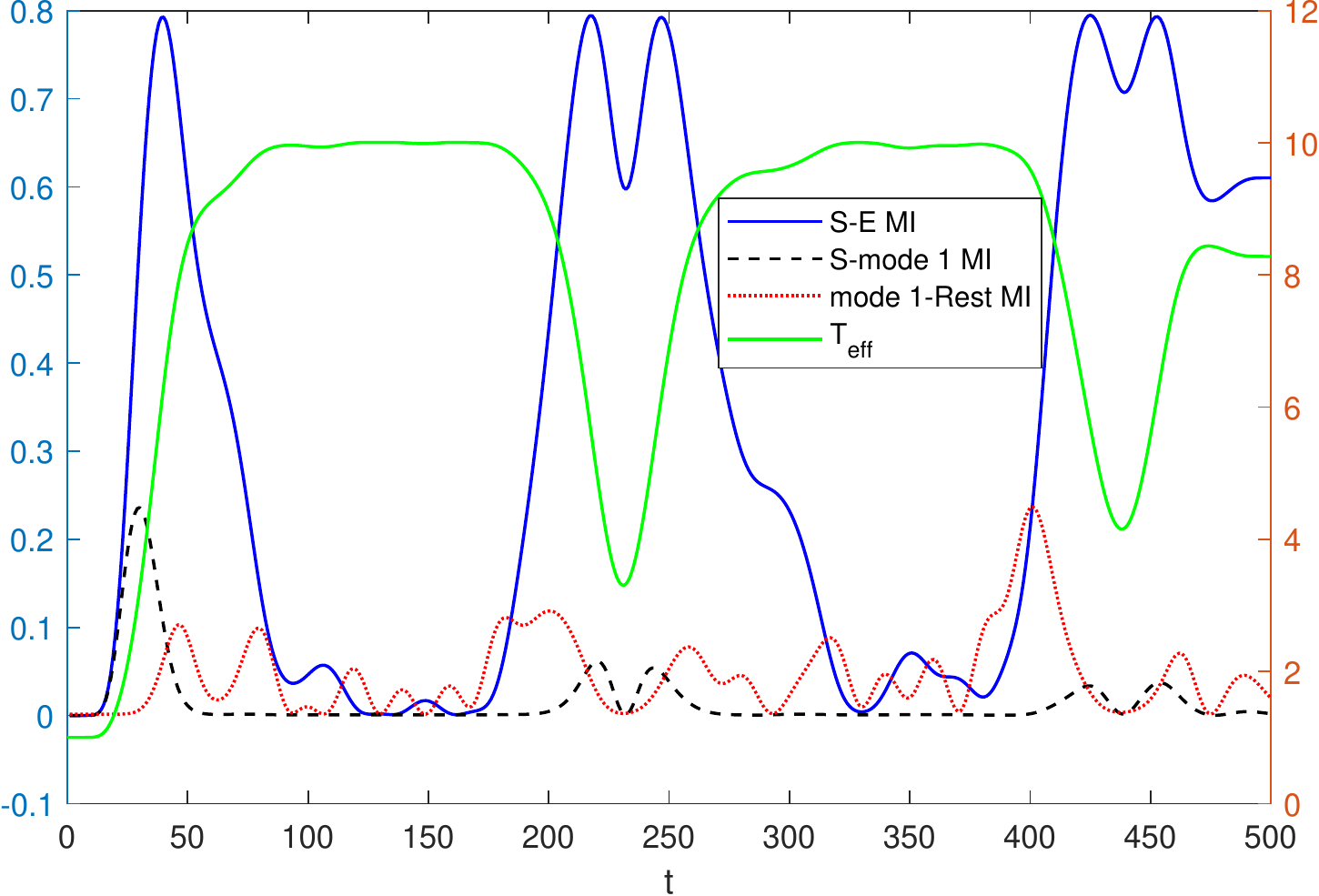}
\caption{Evolution during the interaction between $S$ and $E$ for the oscillator chain. We set $\omega=4$, $N=25$, $T_S=1$, $T_E=10$, $\alpha=\lambda=0.15$, $\delta=0.1\tau$ and $\tau=500$. The green solid line is the effective temperature of the system, corresponding to the right vertical axis, while the other curves are the MI, corresponding to the left vertical axis: the blue solid line is the MI between the $S$ and $E$, black dashed line is the MI between $S$ and mode 1, and the red dotted line is the MI between mode 1 and the rest modes.}
\label{fig3}
\end{center}
\end{figure}

\begin{figure}
\begin{center}
\includegraphics[width=0.45\textwidth]{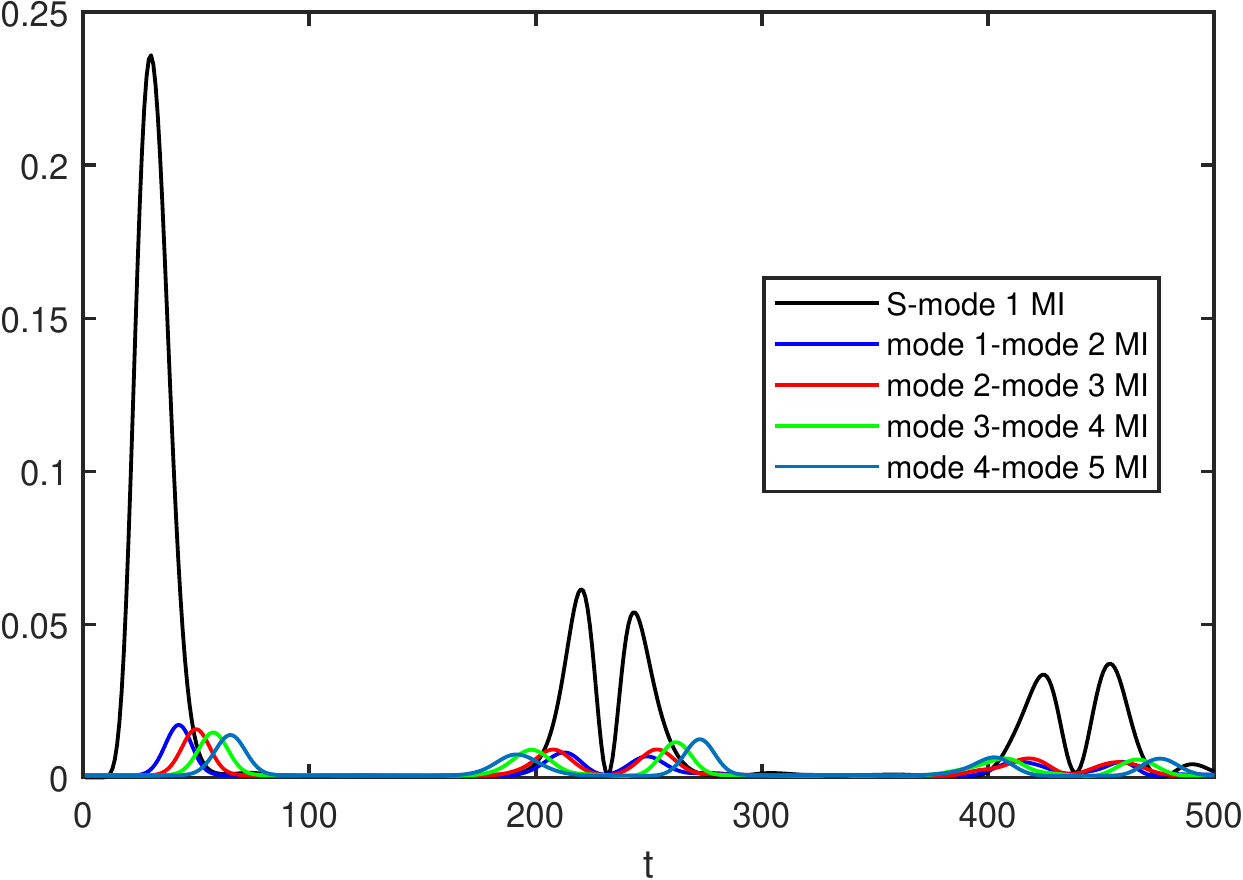}
\caption{Evolution during the interaction between $S$ and $E$ for the oscillator chain. The parameters are set as in the FIG.\ref{fig3}. The black, blue, red, green, and deep blue lines correspond to the MI between System- mode1, mode 1-mode 2, mode 2-mode 3, mode 3-mode 4 and mode 4-mode 5 respectively.}
\label{fig4}
\end{center}
\end{figure}

\begin{figure}
\begin{center}
\includegraphics[width=0.45\textwidth]{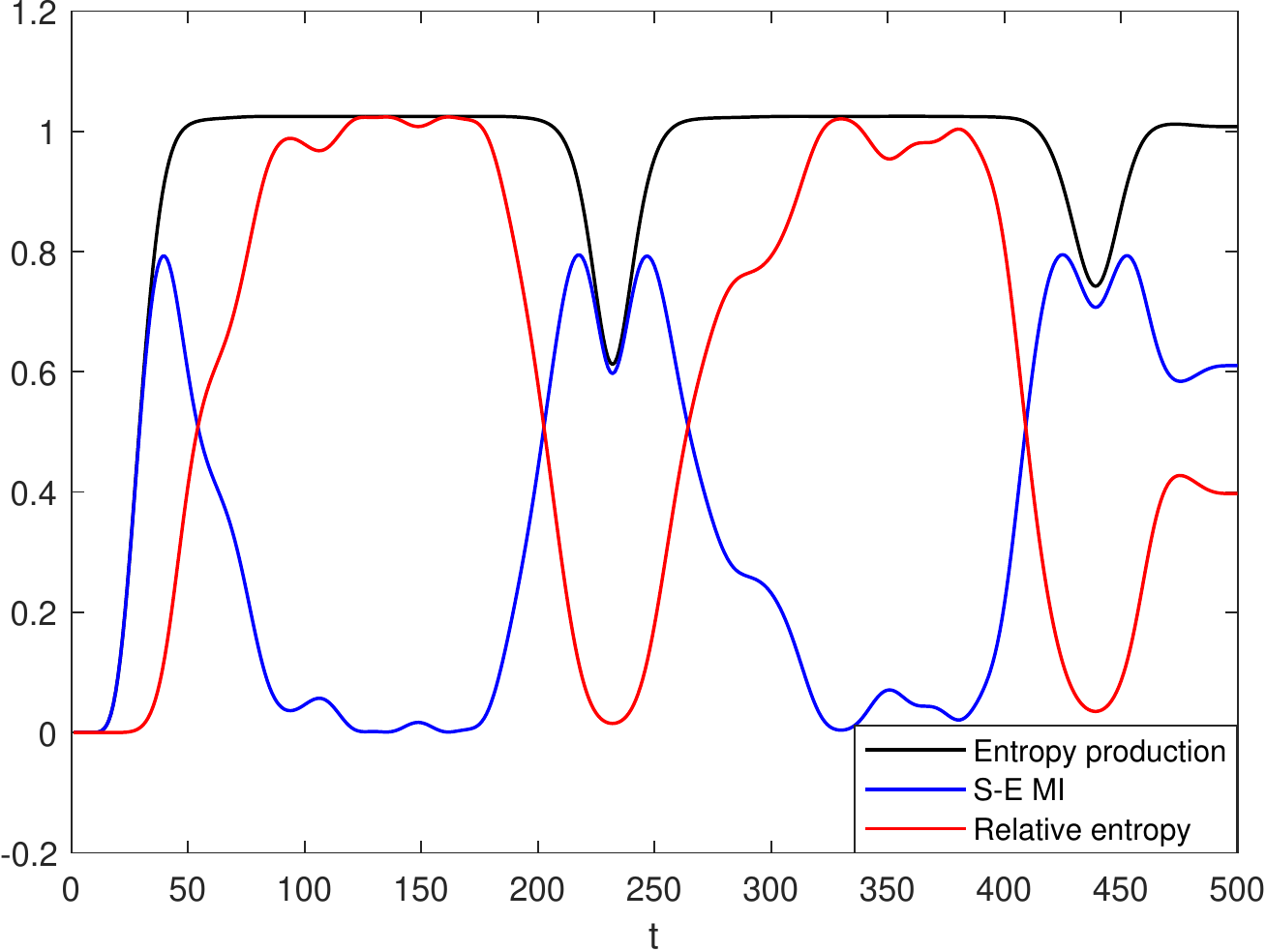}
\caption{Evolution during the interaction between $S$ and $E$ for the oscillator chain. The parameters are set as in the FIG.\ref{fig3}. The black solid line is the entropy production. The blue solid line the MI between $S$ and $E$, and the red solid line is the relative entropy $D(\rho'_E||\rho_E)$.}
\label{fig5}
\end{center}
\end{figure}

In FIG.\ref{fig3}-\ref{fig5} we present some numerical examples of the evolution during the interaction between $S$ and $E$. We set $\omega=4$, $N=25$, $T_S=1$, $T_E=10$, $\alpha=\lambda=0.15$, $\delta=0.1\tau$ and $\tau=500$. In FIG. \ref{fig3} we can find the effective temperature of the system (green line) keeps increasing until it is equivalent to the environment's temperature. The value of the temperature stabilizes for some time, then starts to decrease, reaches a lower point, and then increases again. If we extend the interaction time, we will find the same pattern: the temperature of the system remains stable for a period of time, then experiences an oscillation, after which it returns to stability and waits for the next disturbance. On the other hand, we can also observe that the MI between the system and the environment (blue solid line) shows similar characteristics. The MI between the system and mode 1 also increases at the beginning, but later decreases until the arrival of the disturbance. The increase in MI at the beginning of evolution comes from the fact that we turn on the interaction, while the decrease is due to the internal interactions of the environment that make correlation propagate in the environment. On the other hand, we can find that the MI between mode 1 and the rest is always oscillating.

We take a closer look at the correlation of the environment in FIG.\ref{fig4}, where we consider the correlations of adjacent modes. In addition to the MI between system-mode 1, we also present the evolution of the MI between mode 1-mode 2, mode 2-mode 3, mode 3-mode 4 and mode 4-mode 5. Similar steps can be performed for all oscillators. Due to the periodic boundary condition $\hat{q}_{N+1}=\hat{q}_1$, they are equivalent to the MI of mode 1-mode $N$, mode $N-1$-mode $N$, mode $N-2$-mode $N-1$, mode $N-3$-mode $N-2$, etc. In FIG.\ref{fig4}, we can observe how the correlation propagates through the oscillator chain. Starting from mode 1, the correlation between the adjacent modes increases and then decays one by one. After some time we can find the MI oscillates again, because the correlation from the other side arrives. The correlation makes a complete circle in both directions back to mode 1, causing the oscillation of all the physical quantities, then the correlation between the $S$ and mode 1 decays and the evolution starts a new cycle again.

In FIG.\ref{fig5} we study the evolution of the entropy production, $I(S':E')$ and $D(\rho'_E||\rho_E)$. The behavior is also regular. In the early stage of evolution, entropy production mainly comes from MI, since $S$ and $E$ have just interacted. As the interaction continues, the effective temperature of $S$ increases until it is approximately the same as the temperature of $E$. On the other hand, the MI decreases, meaning the correlation between $S$ and $E$ decays, while relative entropy increase instead, implying the displacement of the environment from equilibrium begins to become significant. When the effective temperature is stable, the relative entropy reaches the highest, while the MI is the lowest. This state will continue for some time, depending on the number of the oscillator in the chain, until the correlation comes back around a full circle and causes a new period. If we set $N=1$, the curves would be exactly like the FIG.\ref{fig2}. For a larger value of $N$, the temperature and other quantities, such as entropy production, $I(S':E')$ or $D(\rho'_E||\rho_E)$, would stay stable for a longer time, because the correlation would take a longer time to travel through the entire oscillator chain.

\emph{Final remarks}.---In summary, the interaction between $S$ and $E$ creats correlations. If the mode in $E$ that interacts with $S$ also interact with other modes in $E$, the correlation will propagate in the $E$, depending on the coupling strength, the number of modes, boundary condition, etc. In the free scalar cavity QFT, the mode that shares the same frequency with the $S$ does not interact with other modes in the cavity QFT, thus the correlation does not propagate to other modes. For the interacting oscillate chain with periodic boundary condition, the $S$ interacts with one mode of the $E$, and the correlation propagates in circles like a kind of ``flow'' so the physical quantities oscillate regularly. Neither MI nor relative entropy has always provided a dominant contribution to the entropy production. In our case the entropy production is mainly from MI in the early stage of the interaction and from relative entropy when the system is in  ``thermal equilibrium''. In fact, if we add more internal interactions to the environment or include more modes of the environment in the interaction Hamiltonian, the evolutionary process of the whole system will become more complicated. There is still much to study about the evolution of the whole system.

Note that the entropy production has also been studied in other bosonic Gaussian systems, such as the quantum Brownian motion (QBM) model for the harmonic potential \cite{Pucci2013,Colla:2021mbe}. Similar oscillations of the thermodynamic quantities are observed and explained as a phenomenon of Poincar\'{e} recurrences. This brings no surprise since we are all considering finite dynamical periodic systems. However, there are two main diferences between our second model and (QBM) model. The first one is that in our second model the bath $B$ \eqref{Hbath} contains the interaction term $\sum_{i=1}^N\alpha \hat{q}_{i}\hat{q}_{i+1}$ between the adjacent oscillators, while in the QBM cases the bath is non-interacting. The second one is we set our interaction \eqref{Hint} to be ``\emph{local}'' because the $S$ interacts with only one oscillator, while the interaction in QBM cases is ``\emph{global}'' since $S$ interacts with all the oscillators. In our case the interaction ``perturbs'' the whole system locally and the correlation is created between $S$ and $\hat{q}_1$, then it passes through the total oscillator chain as in FIG.\ref{fig4} to make circles in both directions due to the interaction term $\sum_{i=1}^N\alpha \hat{q}_{i}\hat{q}_{i+1}$.

Another important motivation to study our second model is that it captures some important features of CFT, and it may shed some light on the AdS/CFT correspondence \cite{Maldacena:1997re,Witten:1998qj}. If we consider the discretized Hamiltonian on a lattice with interaction between adjacent oscillators as \eqref{Hbath}, taking $N\rightarrow \infty$ and massless limit, we can obtain the CFT with central charge $c=1$. Thus once we discrete the CFT, we can apply the GQM to study it. This method has been used to investigate the complexity for thermofield double states \cite{Chapman:2018hou} and the purification of (1+1)-dimensional CFT \cite{Camargo:2020yfv}. In our work we consider the local interaction between the system $S$ and bath $B$, which is similar to the thermalization and revivals after a local quantum quench in CFT \cite{Cardy:2014rqa}, where J. Cardy used the quasiparticle picture to explain the quantum recurrence. The left and right-moving quasiparticles are initially entangled, being emitted at $t=0$ and thereafter moving semiclassically. In a periodic system, an oppositely moving pair of particles will meet again and this should lead to a revival of the initial state. The quasiparticle picture is very similar with the correlation propogating in our work, and in our previous work \cite{Xu2021} we indeed studied the (de-)excition of the QFT in this $S$ and $B$ interaction. Our model and Gaussian techniques may be able to help us to get a better understanding of the time evolution of CFT and the nature of holography. Other models in GQM may also bring new perspectives to the study of AdS/CFT. In recent years quantum information theory has become inter-disciplinary and has been used to study gravity and high energy physics \cite{1409.1231,1411.7041,1503.06237}. There have also been some recent discussions on the connection between quantum gravity and Gaussianity of states \cite{Howl2021}. GQM may also be helpful in these fields. However, we also need to emphasize that it appears quite a step to really gain something new about the full AdS/CFT correspondence. There are still many problems to be studied in this field and we hope to get some interesting results in the near future.

Finally, as we previously suggested, the interaction between $S$ and $E$ can also be regarded as the measurement of $S$ to $E$. Measurements are of great importance in physics. In principle, all measurable observations are also representative of all possible properties of the system. The study of the evolution of joint systems may also help us to understand the nature of the measurements themselves \cite{Xu2022} and facilitate the development of practical models to detect or simulate different systems. We will continue to study this direction in our future work.

\begin{acknowledgments}
Hao Xu thanks Karen V. Hovhannisyan for useful discussion on the numerical method. He also thanks the Natural Science Foundation of the Jiangsu Higher Education Institutions of China (No.20KJD140001) for funding support.
\end{acknowledgments}

\section*{Data Availability Statement}
The datasets generated during and/or analysed during the current study are available from the corresponding author on reasonable request.

\end{document}